\documentclass[a4paper,12pt]{article}
\usepackage[T1]{fontenc}
\usepackage[utf8]{inputenc}
\usepackage[spanish,english]{babel}
\usepackage{authblk}
\usepackage{csquotes}
\usepackage{xpatch}
\usepackage{hyperref}
\hypersetup{hidelinks}
\usepackage[hypcap=false]{caption}
\usepackage[font=small,labelfont=bf]{caption}
\usepackage{amsmath}
\usepackage{amssymb}
\usepackage{amsthm}
\usepackage{mathtools}
\usepackage{mathrsfs}
\usepackage{braket}
\usepackage{bm}
\usepackage{graphicx}
\usepackage{booktabs}
\usepackage{siunitx}
\usepackage{feynmf}

\newcommand{\MeV}{\text{MeV}}

\newcommand{\fm}{\text{fm}}

\newcommand{\mail}[1]{\href{mailto:#1}{\texttt{#1}}}
\newcommand{\keywords}[1]{\textbf{Keywords:} #1}

\title{\textbf{A plausible explanation of $\Upsilon(10860)$}}
\author{R. Bruschini\thanks{\mail{roberto.bruschini@ific.uv.es}}}
\author{P. González\thanks{\mail{pedro.gonzalez@uv.es}}}
\affil{\foreignlanguage{spanish}{Departamento de Física Teórica-IFIC \\
	Universidad de Valencia-CSIC \\
	E-46100 Burjassot(Valencia)}, Spain}
\date{}

\begin{document}

\maketitle

\begin{abstract}
We show that a good description of the $\Upsilon(10860)$ properties, in particular the mass, the $e^+ e^-$ leptonic widths and the $\pi^{+}\pi^{-}\Upsilon(ns)$$\ (n=1,2,3)$ production rates, can be obtained under the assumption that $\Upsilon(10860)$ is a mixing of the conventional $\Upsilon(5s)$ quark model state with the lowest $P-$ wave hybrid state.
\end{abstract}

\begin{center}
\end{center}

\section{Introduction\label{SI}}

The explanation of the large $e^{+}e^{-}\rightarrow\pi^{+}\pi^{-}\Upsilon(  ns)  $ $(  n=1,2,3)  $ widths at $\sqrt{s}=10.866\pm0.002$ GeV near the $\Upsilon(  10860)  $ peak \cite{San16,Che08,PDG18}, about two orders of magnitude larger than those for $\Upsilon(  n^{\prime}s)  \rightarrow\pi^{+}\pi^{-}\Upsilon(1s)  $ $(  n^{\prime}=2,3,4)  ,$ is nowadays a theoretical challenge. This so called \enquote{anomalous} dipion production suggests that either $\Upsilon(  10860)  $ is not the standard $\Upsilon(5s)$ meson, or there is some overlap of $\Upsilon(5s)$ with a non standard resonance close by, or there are some dynamical effects with much bigger influence for $\Upsilon(5s)$ than for $\Upsilon(  n^{\prime}s)  $ $( n^{\prime}=2,3,4)  $ \cite{Ols17}. Regarding the first option a tetraquark interpretation of $\Upsilon(  10860)  $ was used in reference \cite{Ali10}. By assuming a nonresonant part of the amplitude with the experimentally required order of magnitude the authors showed that the consideration of resonant terms from intermediate $f\Upsilon(  ns)  $ states with $f$ standing for $f_{0}(  500)  ,$ $f_{0}(  980)  $ and $f_{2}(  1270)  $ allowed for a fit of the decay distributions of $\Upsilon(  10860)  \rightarrow\pi^{+}\pi^{-}\Upsilon( ns)  $ $(  n=1,2,3)  .$ However no theoretical justification for the order of magnitude of the nonresonant part was given. Concerning the second and third options we shall try to show next that they may be related in such a way to provide a plausible explanation of $\Upsilon( 10860)  .$ For this purpose we develop in Section~\ref{SII} a standard description of $0^{-}(  1^{--})  $ bottomonium states from a conventional quark potential model. In Section~\ref{SIII} dipion transitions from $\Upsilon(5s)$ are studied within the QCD Multipole Expansion framework through the implementation of intermediate hybrid states. The calculated widths for these processes suggest that a detailed explanation of data is feasible. Finally in Section~\ref{SIV} possible interpretations of $\Upsilon(10860)$ deriving from this explanation are discussed.

\section{Quark potential model\label{SII}}

Our starting point will be the simplest yet realistic non relativistic quark model description of bottomonium $(  b \overline{b})  $ provided by a Cornell like potential \cite{Eic80}
\begin{equation}
V_{C}(  r)  =\sigma r-\frac{\zeta}{r} \label{Cor}
\end{equation}
where $r$ is the $b-\overline{b}$ distance and the parameters $\sigma$ and $\zeta$ stand for the string tension and the chromoelectric coulomb strength respectively. This form for the static potential has been justified from quenched lattice QCD calculations, see for instance \cite{Bal01}. It should be kept in mind that in the spirit of the nonrelativistic quark model calculations $\sigma$ and $\zeta$ have to be considered as effective parameters through which some non considered corrections to the potential may be implicitly taken into account. We shall fix the Coulomb strength to $\zeta=100\,\MeV\,\fm$ corresponding to a strong quark-gluon coupling $\alpha _{s}=\frac{3\zeta}{4\hbar}\simeq0.38$ in agreement with the value derived from QCD from the hyperfine splitting of $1p$ states in bottomonium \cite{Ynd95}. As for $\sigma$ we shall choose its value altogether with the quark mass value $m_{b}$ to get a good fit to the masses of $0^{-}(  1^{--})  $ spin triplet bottomonium states$.$ Thus, for $\sigma=873$ MeV/fm and $m_{b}=4793$ MeV a nice description of the spectral masses is obtained, as shown in Table~\ref{Tabmass}.

\begin{table}
\centering
\begin{tabular}{ccc}
\toprule
$nl\text{ States}$ & $M_{nl}\text{ (MeV)}$ & $M_{PDG}\text{ (MeV)}$\\
\midrule
$1s$ & $9463$ & $9460.30\pm0.26$\\
$2s$ & $10023$ & $10023.26\pm0.31$\\
$1d$ & $10169$ & $10163.7\pm1.4$\\
$3s$ & $10358$ & $10355.2\pm0.5$\\
$2d$ & $10455$ & \\
$4s$ & $10628$ & $10579.4\pm1.2$\\
$3d$ & $10703$ & \\
$5s$ & $10865$ & $10889.9_{-2.6}^{+3.2}$\\
$4d$ & $10926$ & \\
$6s$ & $11081$ & $10992.9_{-3.1}^{+10.0}$ \\
\bottomrule
\end{tabular}
\caption[Calculated $0^{-}(  1^{--})  $ bottomonium masses.]{Calculated $0^{-}(  1^{--})  $ bottomonium masses, $M_{nl}$, from $V_{C}(  r)  $ with $\sigma=873$ MeV/fm, $\zeta=100\,\MeV\,\fm$ and $m_{b}=4793$ MeV. The spectral notation $nl$ where $n$ $( l)  $ indicates the principal (orbital angular momentum) number has been used for the states. For the $ns$ and the $1d$ states the masses of the closest experimental $\Upsilon$ resonances from \cite{PDG18}, $M_{PDG}$, are quoted for comparison.}
\label{Tabmass}
\end{table}

Some comments are in order. First, the significant discrepancy between the calculated mass of the $4s$ state, $10628$ MeV, and the experimental measured mass at $10579.4$ MeV may be indicating mixing of the $4s$ and $3d$ states. So, the measured resonance would have a dominant $4s$ component, whereas a not yet discovered resonance at about $10750$ MeV would have a dominant $3d$ component. Second, the discrepancy between the calculated mass of the $6s$ state, $11081$ MeV, and the experimental measured mass at $10992.9$ MeV indicates the need for including the effect of the first $S-$ wave open bottom meson-meson channel $B\overline{B}_{1}$ in the potential when crossing the $B\overline{B}_{1}$ threshold, see \cite{Gon14}. Third, the natural assignment of $\Upsilon(  10860)$ is to the $\Upsilon(5s)$ state since the corresponding peak observed in the $e^{+}e^{-}\rightarrow b\overline{b}$ cross section is about the $5s$ calculated energy \cite{San16}. It should be kept in mind though that some mixing with the $4d$ state can also be expected.

As for the error in the calculated masses the effectiveness of the parameters makes difficult to quantify it. We should expect for instance relativistic effects to be more important for the low lying states. Then, having chosen the values of the parameters as to fit these states may produce a non physical mass shift for the high lying ones. In this sense the 25 MeV difference between the calculated mass of $\Upsilon(5s)$ and the quoted value for the mass of $\Upsilon(10860)$ might be taken as a very rough estimate of the error.

It is easy to check that the calculated $^{3}S_{1}$ states provide a very good description of the measured ratios $\frac{\Gamma(  \Upsilon( n_{1}s)  \rightarrow e^{+}e^{-})  }{\Gamma(  \Upsilon( n_{2}s)  \rightarrow e^{+}e^{-})  }$ and the correct order of magnitude for radiative transitions to $^{3}P_{1}$ states, for example
\[
\frac{\Gamma(  \Upsilon(  3s)  \rightarrow\chi_{b1}( 1p)\gamma )  }{ \Gamma(  \Upsilon(  3s) \rightarrow\chi_{b1}(  1p)  \gamma) _\text{Exp} }=\frac{4}{20\pm10}
\]
and
\[
\frac{\Gamma(  \Upsilon(  3s) \rightarrow\chi_{b1}(  2p)\gamma  )  }{\Gamma( \Upsilon(  3s)  \rightarrow\chi_{b1}(  2p) \gamma)_\text{Exp}}=\frac{2.7}{2.7\pm0.3}.
\]

\section{Dipion transitions $\Upsilon(  n_{i}s)  \rightarrow\pi ^{+}\pi^{-}\Upsilon(  n_{f}s)  $ \label{SIII}}

Let us now center on the dipion transitions between $0^{-}( 1^{--})  $ $s$ states: $\Upsilon(  n_{i}s)  \rightarrow \pi^{+}\pi^{-}\Upsilon(  n_{f}s) $. In QCD these processes involve the emission of two gluons and the conversion of gluons into pions. As far as the heavy quark system moves slowly and its size is small compared to the$\ $pion system a non relativistic treatment based on the QCD Multipole Expansion (QCDME) makes sense, see \cite{Kua09} and references therein. Then the transition rate, dominated by double electric dipole transitions, can be expressed as \cite{Kua81}
\begin{equation}
\Gamma(  \Upsilon(  n_{i}s)  \rightarrow\pi^{+}\pi^{-}\Upsilon(  n_{f}s)  )  =CG\lvert F_{n_{i}n_{f}}^{1}\rvert ^{2} \label{tranrate}
\end{equation}
where $C$ is a constant whose value can be fixed from a fit to data (see below in this section), $G$ is the phase space factor
\begin{equation}
G=\frac{3}{4}\frac{M_{n_{f}s}}{M_{n_{i}s}}\frac{\pi^{3}}{\hbar^{4}}\int \mathrm{d}M_{\pi\pi}^{2} K\sqrt{1-\frac{4m_{\pi}^{2}}{M_{\pi\pi}^{2}}}(  M_{\pi\pi}^{2}-2m_{\pi }^{2})  ^{2}
\label{phasespace}
\end{equation}
with $M_{\pi\pi}$ the dipion invariant mass, and
\begin{equation}
K=\frac{\sqrt{(M_{n_{i}s}+M_{n_{f}s})^{2}-M_{\pi\pi}^{2}}\sqrt{(M_{n_{i} s}-M_{n_{f}s})^{2}-M_{\pi\pi}^{2}}}{2M_{n_{i}s}} \label{key}
\end{equation}
is the recoil momentum of $\Upsilon(  n_{f}s)  $ in the rest frame of $\Upsilon(  n_{i}s) $. The transition matrix element$F_{n_{i}n_{f}}^{1}$ is given by
\begin{equation}
F_{n_{i}n_{f}}^{1}=\sum_{n_{hyb}}\frac{\int \mathrm{d}r \, r^{2} R_{n_{i}s}(  r)  r R_{n_{hyb}p}( r) \int \mathrm{d}r'\,r'^{2} R_{n_{hyb}p}(  r') r' R_{n_{f}s}(  r')  }{M_{n_{i}s}-\mathcal{M}_{n_{hyb}p}}
\label{F}
\end{equation}
where $R$ stands for the radial wave function and the sum runs over a complete set of color singlet intermediate states of angular momentum $1$, each of them containing a $b\overline{b}$ color octet. We identify these intermediate states as hybrids $(  (  b\overline{b})  ^{8}+gluon)  $ denoted by their principal ($ n_{hyb}$) and orbital angular momentum ($l=1$) quantum numbers.

Potentials for hybrid states have been derived in quenched lattice QCD \cite{Jug99} and parametrized in reference \cite{Bra14}. We shall assume that the dominant contribution to the sum in~\eqref{F} comes from the hybrid states with orbital angular momentum $1$ corresponding to the deepest hybrid potential called $V_{\Pi_{u}}$. The lowest energy hybrid is indeed the $1p$ state of $V_{\Pi_{u}}$. At short and intermediate distances this potential has been parametrized as
\[
V_{\Pi_{u}}(  r)  =\biggl(\frac{0.24}{r_{0}^{3}}r^{2}+\frac{0.11} {r}\biggr)\hbar+E_{\Pi_{u}}
\]
where $r_{0} \simeq 0.5\,\fm$ and $E_{\Pi_{u}}$ is an additive constant, while at large $r$ it reads
\[
V_{\Pi_{u}}(  r)  \rightarrow\sigma r\sqrt{1+\frac{11\pi\hbar }{6\sigma r^{2}}}+E_{0}
\]
where $E_{0}$ is another additive constant so that $E_{\Pi_{u}}-E_{0}\simeq\frac{2.5\,\MeV\,\fm}{r_{0}}$, as to ensure that the two parametrizations connect smoothly. It is important to realize that $E_{\Pi_{u}}$ corresponds to the energy of the ground state $1^{+-}$ gluelump (formed by a gluon bound to a $b\overline{b}$ color octet located at the origin). This energy has been estimated to be between $740$ MeV and $1040$ MeV \cite{BaPi04}.

\begin{figure}
\centering
\includegraphics[width=3.659in]{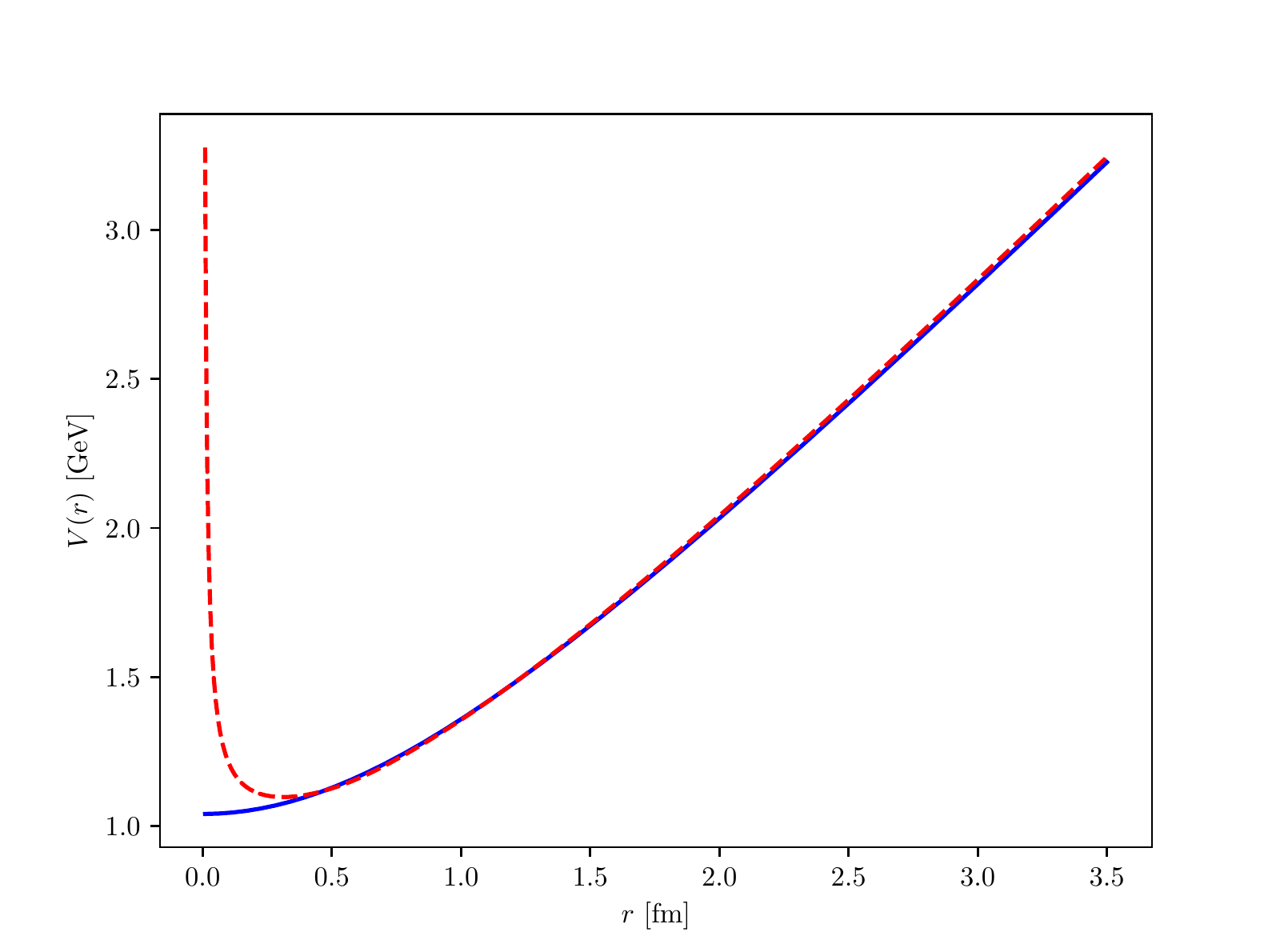}
\caption[\text{\emph{Hybrid}} potential versus vibrational potential]{\text{\emph{Hybrid}} potential $V_{\Pi_{u}}( r)$ with $r_0 = 0.51\,\fm$, $E_{0}=31.4\,\MeV$ and $E_{\Pi_{u}}=992.3\,\MeV$ (dashed line) versus vibrational potential $V_{vib}(r)$ (solid line).}
\label{Fig1}
\end{figure}

This parametrization of $V_{\Pi_{u}}$ resembles the form of the deepest vibrational string potential derived in reference \cite{Gil77}
\begin{equation}
V_{vib}(r)=\sigma r\sqrt{1+\frac{2\pi\hbar}{\sigma r^{2}}}
\label{vib}
\end{equation}
except for its short range behavior since $V_{\Pi_{u}}$ becomes a repulsive Coulomb potential (with a reduced strength as compared to $\zeta$) instead of the constant potential resulting from $V_{vib}(r\rightarrow0)$. This is illustrated in Figure~\ref{Fig1} where $V_{vib}(r)$ has been drawn versus $V_{\Pi_{u}}(  r)  $ with $E_{\Pi_{u}}\simeq990$ MeV and $E_{0}\simeq30$ MeV. 

As a matter of fact the reduced short range repulsion has little effect on the masses of the intermediate states we are interested in, and on the calculation of $F_{n_{i}n_{f}}^{1},$ so that one can safely use the simpler compact expression of $V_{vib}(r)$ instead of $V_{\Pi_{u}}(  r)  .$ For the sake of simplicity and for an easy comparison to other vibrational potentials used in the literature within the QCDME framework we shall use $V_{vib}(r)$ henceforth.

It turns out that the mass of the lowest hybrid state, $\mathcal{M}_{1p}=10888$ MeV, is pretty close to the calculated mass of the $5s$ state, $M_{5s}=10865$ MeV (the masses for the higher hybrid states are $\mathcal{M}_{2p}=11082$ MeV, $\mathcal{M}_{3p}=11267$ MeV...) This gives rise to an enhancement of the amplitudes~\eqref{F} for $\Upsilon(5s)$ as compared to $\Upsilon(  n_{i}s)  $ $(  n_{i}<5)  $. More precisely, by making use of a sufficient number of hybrid states (equal or greater than $10$) as to assure convergence of the sum in~\eqref{F} and fixing the constant $C=6.53\times10^{-5}$ to get the experimental $\Upsilon(  2s)  \rightarrow\pi^{+}\pi^{-}\Upsilon( 1s)  $ width we can reproduce nicely the order of magnitude for all the $\Upsilon(  n_{i}s)  \rightarrow\pi^{+}\pi^{-}\Upsilon( n_{f}s)  $ widths with $n_{i}\leq5,$ with the exception of $\Upsilon(  5s)  \rightarrow\pi^{+}\pi^{-}\Upsilon( 3s)  $ as can be checked in Table~\ref{Dipion}.

\begin{table}
\centering
\begin{tabular}{rrrr}
\toprule
$\text{Process}\ \ \ \ \ \ \ \ $ & $\Gamma$(keV) & $\mathcal{B}%
_{QCDME}\ \ \ \ \  $ & $ \mathcal{B}_{PDG}\ \ \ \ \ \ \ \ \ \ \ $\\
\midrule
$\Upsilon(  3s)  \rightarrow\pi^{+}\pi^{-}\Upsilon(
1s)  $ & $\ \ 0.936$ & $(  4.6\pm0.4)  \times10^{-2}$ &
$(  4.37\pm0.08)  \times10^{-2}$\\
$\Upsilon(  3s)  \rightarrow\pi^{+}\pi^{-}\Upsilon(
2s)  $ & $\ \ 0.575$ & $(  3.0\pm0.3)  \times10^{-2}$ &
$(  2.82\pm0.18)  \times10^{-2}$\\
$\Upsilon(  4s)  \rightarrow\pi^{+}\pi^{-}\Upsilon(
1s)  $ & $\ \ 6.932$ & $(  3.4\pm0.4)  \times10^{-4}$ &
$\ \ \ (  8.2\pm0.4)  \times10^{-5}$\\
$\Upsilon(  4s)  \rightarrow\pi^{+}\pi^{-}\Upsilon(
2s)  $ & $\ \ 3.995$ & $(  1.9\pm0.2)  \times10^{-4}$ &
$\ \ \ (  8.2\pm0.8)  \times10^{-5}$\\
$\Upsilon(  5s)  \rightarrow\pi^{+}\pi^{-}\Upsilon(
1s)  $ & $\ \ 655.7$ & $(  1.2\pm0.2)  \times10^{-2}$ &
$\ \ \ (  5.3\pm0.6)  \times10^{-3}$\\
$\Upsilon(  5s)  \rightarrow\pi^{+}\pi^{-}\Upsilon(
2s)  $ & $\ \ 115.9$ & $(  2.3\pm0.3)  \times10^{-3}$ &
$\ \ \ (  7.8\pm1.3)  \times10^{-3}$\\
$\Upsilon(  5s)  \rightarrow\pi^{+}\pi^{-}\Upsilon(
3s)  $ & $\ \ \ \ 20.6$ & $\ \ \ (  4.0_{-0.6}^{+0.7})
\times10^{-4}$ & $\ \ \ \ \ \ (  4.8_{-1.7}^{+1.9})  \times10^{-3}$\\
\bottomrule
\end{tabular}
\caption[Calculated widths and branching fractions for dipion transitions between $\Upsilon(ns)  $ states.]{Calculated widths and branching fractions $\mathcal{B}_{QCDME}$ for dipion transitions between $\Upsilon(  ns)  $ states within the QCDME framework. The errors in $\mathcal{B}_{QCDME}$ come from the errors in the experimental values of the total widths. Experimental branching fractions from \cite{PDG18}, $\mathcal{B}_{PDG},$ are quoted for comparison.}
\label{Dipion}
\end{table}

A look in detail at this table shows that the calculated widths from $\Upsilon(  3s)  $ are in perfect agreement with data; from $\Upsilon(  4s)  $ the calculated widths are bigger (by at most a factor $4)$ as should be expected if the experimental resonance has some $3d$ mixing (for the suppression of dipion decays from $d$ states see \cite{Mox88}). Regarding $\Upsilon(5s)$ the calculated dipion widths to $\Upsilon(  1s)  $ and $\Upsilon(  2s)  $ have the correct order of magnitude differing from data by at most a factor $3$ whereas in the decay to $\Upsilon(  3s)  $ the calculated width is one order of magnitude lower than data.

It should be kept in mind though that there are several sources of error in the calculated widths. First, in the fixing of $C:$ as we rely on the PDG average value of the $\Upsilon(  2s)  \rightarrow\pi^{+}\pi ^{-}\Upsilon(  1s)  $ width to fix it we estimate a small $2\%$ error; this can be taken as a minimum possible error since the experimental dispersion of data is much bigger. Second, in the truncated series of intermediate states; by comparing the calculated widths with different number of terms we estimate this error to be another $2\%.$ Third, in the use of the QCDME because of its expected lost of accuracy when increasing $n_{i}$ due to the higher size of the initial state; this can not be trustly estimated. Nonetheless, the good values obtained for the decays of $\Upsilon(  3s)  $ and $\Upsilon(  4s)  $ make us confident that the calculational error in the $\Upsilon(5s)$ case does not affect the calculated order of magnitude.

\begin{figure}
\centering
\includegraphics[width=3.659in]{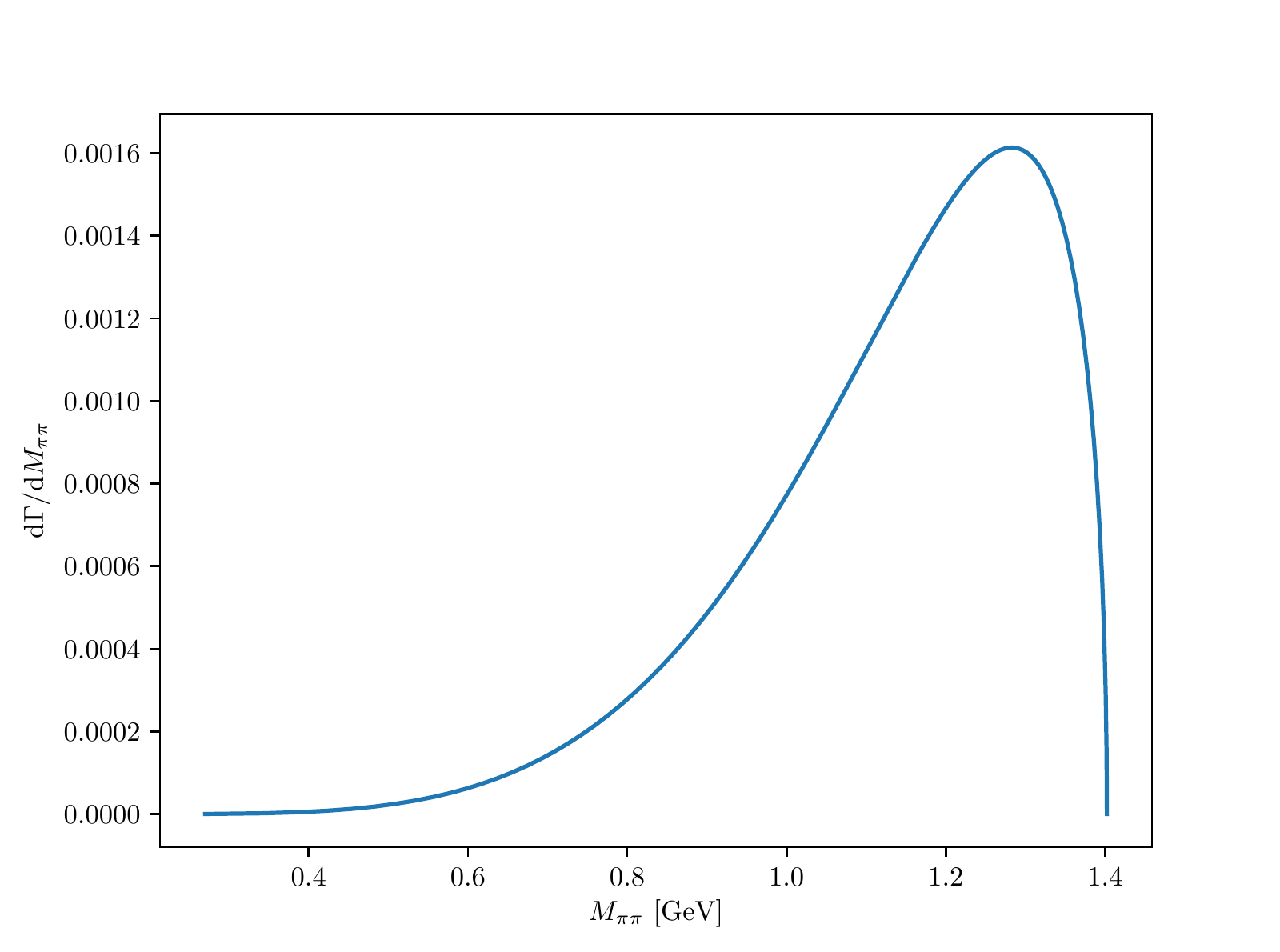}
\caption[Dipion invariant mass distribution for $\Upsilon(5s)  \rightarrow\pi^{+}\pi^{-}\Upsilon(  1s) $.]{Calculated dipion invariant mass distribution for $\Upsilon(5s)  \rightarrow\pi^{+}\pi^{-}\Upsilon(  1s) $.}
\label{Fig2}
\end{figure}

A more precise interpretation of the results in Table~\ref{Dipion} requires an analysis of the dipion invariant mass distribution $\frac{d\Gamma}{dM_{\pi\pi}}$ in the way it was carried out for instance in reference \cite{Ali10}. Our calculated $\frac{d\Gamma}{dM_{\pi\pi}}$ from~\eqref{tranrate}, plotted in Figure~\ref{Fig2}, should be identified as the nonresonant part of the amplitude (see for comparison Figure 2 in \cite{Ali10}). Thus, our model provides a physical justification to the educated guess done in \cite{Ali10} for the $S-$ wave nonresonant amplitude (as we do not consider any mixing of $\Upsilon(5s)$ with $\Upsilon(  4d)  $ we have no $D-$ term).

Regarding additional contributions to the amplitude a look at the experimental representations of $\frac{d\Gamma}{dM_{\pi\pi}}$ versus $M_{\pi\pi}$ for the dipion decay $\Upsilon(  5s)  \rightarrow\pi^{+}\pi^{-}\Upsilon( n_{f}s)  $, see \cite{Ali10} and \cite{Gar15}, shows clearly enhancements suggesting the presence of resonant terms where the two pions are produced via a $0^{+}(  0^{++},2^{++})  $ resonance. In the QCDME framework these would correspond to contributions to the amplitude where the conversion of the two gluons to two pions takes place through a $0^{+}( 0^{++},2^{++})  $ resonance. Following reference \cite{Bro75} a $f_{0}(  500)  $ contribution would contain, up to a dimensional constant, a factor $\frac{M_{\pi\pi}^{2}-m_{\pi}^{2}}{M_{f_{0}( 500)  }^{2}-M_{\pi\pi}^{2}}$ substituting the factor $M_{\pi\pi}^{2}-2m_{\pi}^{2}$ in~\eqref{phasespace}. As the numerator peaks at large $M_{\pi\pi}$ (remember that $M_{\pi\pi}\leq M(  5s)  -M( n_{f}s)  $), and the denominator at $M_{\pi\pi}=M_{f_{0}( 500)  }$ it is clear that the closer to $0$ the difference between $M_{f_{0}(  500)  }$ and $(  M(  5s)  -M( n_{f}s)  )  $ the more important this contribution. As the mass of $f_{0}(  500)  $ is about $(  M(  5s)  -M( 3s)  )  $ we expect it to be dominant for $\Upsilon( 5s)  \rightarrow\pi^{+}\pi^{-}\Upsilon(  3s)  $ and subdominant for $\Upsilon(  5s)  \rightarrow\pi^{+}\pi^{-}\Upsilon(  n_{f}s)  $ with $(  n_{f}=1,2) $. This provides a qualitative explanation of the order of magnitude discrepancy between the calculated nonresonant width and data in the $\Upsilon( 5s)  \rightarrow\pi^{+}\pi^{-}\Upsilon(  3s)  $ case. (As for the estimation of other resonant contributions like the ones coming from intermediate $Z_{b}^{\pm}\pi^{\mp}$ states a theoretical calculational scheme has not been completely developed yet).

\section{Nature of $\Upsilon( 10860)  $\label{SIV}}

The previous results on dipion decays point out to a possible interpretation of $\Upsilon(  10860)  $ as the standard $\Upsilon(5s)  $ state. Further support to this interpretation seems to be provided by the leptonic width ratios calculated from $\Upsilon(5s)$ as
\begin{equation}
\mathcal{R}(n) \equiv \frac{\Gamma(\Upsilon(  5s)  \rightarrow e^{+}e^{-})
}{\Gamma(\Upsilon(  ns)  \rightarrow e^{+}e^{-})} = \frac{\lvert R_{\Upsilon(  5s)  }(  0)  \rvert ^{2}}{\lvert R_{\Upsilon(  ns)  }(  0)  \rvert ^{2}}\frac{M_{\Upsilon(  ns)  }^{2}}{M_{\Upsilon(  5s)  }^{2}}
\label{lepwidth}
\end{equation}
As can be checked from Table~\ref{e+e-} the resuls for $n=1,2,3$ (for $n=4$ mixing with the $3d$ state should be taken into account) are in perfect agreement with experimental ratios $\frac{\Gamma(\Upsilon(10860)\rightarrow e^{+}e^{-})_\text{Exp}}{\Gamma(\Upsilon( ns) \rightarrow e^{+}e^{-})_\text{Exp}}$ . Notice that this also precludes a significant mixing of the $\Upsilon(5s)$ with the $\Upsilon(4d)$ state.

\begin{table}
\centering
\begin{tabular}{lll}
\toprule
$n$ & $\mathcal{R}(n)$ & $\mathcal{R}(n)_\text{Exp}$\\
\midrule
$1$ & $0.19$ & $0.23\pm0.05$\\
$2$ & $0.51$ & $0.52\pm11$\\
$3$ & $0.71$ & $0.70\pm0.16$ \\
\bottomrule
\end{tabular}
\caption[Calculated leptonic width ratios]{Calculated leptonic width ratios $\mathcal{R}(n)$ from $\Upsilon(5s)$, compared to experimental values $\mathcal{R}(n)_\text{Exp}$ from \cite{PDG18}.}
\label{e+e-}
\end{table}

However, this interpretation can not be maintained when dipion decays $\Upsilon(5s)\rightarrow\pi^+\pi^- h_b(np)$ are examined. From the experimental point of view the production rates of $\Upsilon(10860)\rightarrow\pi^+\pi^- h_b(np)$ and  $\Upsilon(10860)\rightarrow\pi^+\pi^- \Upsilon(n_f s)$ are of the same order of magnitude \cite{Ada12}. From the theoretical side the QCDME has no predictive power for these $E1-M1$ transitions (the only available data for $n_i<5$, $\Gamma(\Upsilon(3s)\rightarrow\pi^+\pi^- h_b(1p))<(2.4\pm0.2)\times 10^{-3}\,\MeV$, does not allow for the fixing of the unknown constants). Nonetheless a simplified order of magnitude estimate can be obtained by approximating hadronic transition rates by gluon emission rates. Following reference \cite{Kua81} we can calculate the ratio
\[
\frac{\Gamma(\Upsilon(5s)\rightarrow \pi^+ \pi^- h_b(1p))}{\Gamma(\Upsilon(3s)\rightarrow \pi^+ \pi^- h_b(1p))} \approx \frac{\Gamma(\Upsilon(5s)\rightarrow g g \, h_b(1p))}{\Gamma(\Upsilon(3s)\rightarrow g g \, h_b(1p))} = \frac{(M_{5s}-M_{1p})^7}{(M_{3s}-M_{1p})^7} \frac{\lvert \mathfrak{g}_{5,1}\rvert^2}{\lvert \mathfrak{g}_{3,1}\rvert^2}
\]
where (notice that the potential $V_{\Pi_u}(r)$ does not have $S-$ wave hybrid states \cite{Bra14})
\[
\mathfrak{g}_{n_i,1} \equiv \sum_{n_{hyb}} \frac{\int \mathrm{d}r\,r^2 R_{n_i s}(r) r R_{n_{hyb}p}(r)\int \mathrm{d}r'\,r'^2 R_{n_{hyb}p}(r') R_{h_b(1p)}(r')}{M_{n_i s}-\mathcal{M}_{n_{hyb}p}}
\]
In our spin independent quark potential model $V_C(r)$ the spin singlet $h_b(1p)$ and the spin triplet $\chi_{b1}(1p)$ are degenerate. Then using $R_{h_b(1p)}(r)=R_{\chi_{b1}(1p)}(r)$ we get
\[
\frac{\Gamma(\Upsilon(5s)\rightarrow \pi^+ \pi^- h_b(1p))}{\Gamma(\Upsilon(3s)\rightarrow \pi^+ \pi^- h_b(1p))} \approx 1.1\times 10^2
\]
This theoretical ratio is at least two order of magnitude smaller than data
\[
\frac{\Gamma(\Upsilon(10860)\rightarrow \pi^+ \pi^- h_b(1p))_\text{Exp}}{\Gamma(\Upsilon(3s)\rightarrow \pi^+ \pi^- h_b(1p))_\text{Exp}} > 7.3 \times 10^4
 \]
making the interpretation of $\Upsilon(10860)$ as the standard $\Upsilon(5s)$ state untenable.

The simplest possible alternative is to interpret $\Upsilon(10860)$ as a result of the mixing of $\Upsilon(5s)$ with the first hybrid that we shall call henceforth $H_b(1p)$.  This seems quite natural for the $\Upsilon(5s)$ and the $H_b(1p)$ masses are both close to the measured mass of $\Upsilon(10860)  $ (for the sake of simplicity we do not include any possible mixing with the $\Upsilon(4d)$ state). We may then write
\begin{equation}
\ket{\Upsilon(  10860)} \approx\cos \theta\ket{\Upsilon(  5s)} +\sin\theta \ket{H_b(1p)}
\label{mixing}
\end{equation}
Let us first emphasize that the good description of the $\pi^{+}\pi^{-}\Upsilon(  n_{f}s)  $ decays and the leptonic width ratios obtained from $\Upsilon(5s)$ points out to a small mixing angle. Then, following reference \cite{Bur03} we write
\begin{equation}
\sin\theta\approx\frac{\braket{\Upsilon(  5s)  \rvert \delta\mathcal{H}\lvert H_b(1p)} } {M_{5s}-\mathcal{M}_{H_b(1p)  }}
\label{sin}
\end{equation}
where $\delta\mathcal{H}$ is proportional to the $E1$ transition operator since $\Upsilon(5s)$ and $H_b(1p)  $ have orbital angular momentum $0$ and $1$ respectively (notice that in reference \cite{Bur03} the mixing of $\Upsilon(1s)$ with a different hybrid is considered). Hence we can rewrite the mixing as
\[
\sin\theta\approx A\frac{\int \mathrm{d}r\,r^{2}R_{5s}(  r)  r R_{H_b(1p)  }(  r)  }{M_{5s}-\mathcal{M}_{H_b(1p)}}= A(2\times10^{-4}\fm/\MeV)
\]
where the proportionality constant $A$ has units MeV/fm. By defining $A\equiv a\sigma$ where $\sigma=873 \,\MeV/\fm$ stands for the confining strength for standard as well as hybrid states we get
\[
\sin\theta\approx 0.17 a
\]
being $a$ a dimensionless constant.

As this mixing allows for $H_b(1p)  $ to decay to $e^{+}e^{-}$ through its coupling to $\Upsilon(5s)$ we can estimate
\[
\Gamma(  H_b(1p)  \rightarrow e^{+}e^{-})
\approx0.03a^{2}\Gamma(  \Upsilon(  5s)  \rightarrow
e^{+}e^{-})
\]
Then, taking into account that the calculated leptonic width ratios from $\Upsilon(5s)$ leave very small room for corrections we may reasonably assume $a^{2}$ to be at most of order $1.$ This corresponds to a mixing of at most a few percent. Thus, the good description of the $\pi^{+}\pi^{-}\Upsilon(  n_{f}s)  $ decays previously obtained from
$\Upsilon(5s)$ is also preserved if we reasonably assume, from Heavy Quark Spin Symmetry, that $H_b(1p)\rightarrow\pi^+\pi^-\Upsilon(n_f s)$ is somewhat suppressed against $H_b(1p)\rightarrow\pi^+\pi^-h_b(n p)$. In this regard let us remind that $(S_{b\bar{b}})_{H_b(1p)}=(S_{b\bar{b}})_{h_b(1p)}=0 \ne (S_{b\bar{b}})_{\Upsilon(n s)} = 1$.

The remaining issue has to do with the dipion decays $\Upsilon(10860)  \rightarrow\pi^{+}\pi^{-}h_{b}(  np)$. According to our discussion above, the $\Upsilon(5s)\rightarrow\pi^+\pi^- h_b(1p)$ decay should give a small contribution. So, we should have
\[
\Gamma(\Upsilon(10860)\rightarrow\pi^+\pi^- h_b(np)) \approx \sin^2\theta \, \Gamma(H_b(1p)\rightarrow\pi^+\pi^- h_b(np))
\]
Then, using the experimental widths $\Gamma(\Upsilon(10860)\rightarrow\pi^+\pi^- h_b(1p))_\text{Exp}=(1.8\pm0.9)\times 10^{-1}\MeV$, $\Gamma(\Upsilon(10860)\rightarrow\pi^+\pi^- h_b(2p))_\text{Exp}=(2.9\pm1.5)\times 10^{-1}\MeV$ and $\sin^2\theta \le 0.1$ we can predict
\[
\begin{aligned}
\Gamma(H_b(1p)\rightarrow\pi^+\pi^- h_b(1p)) &\ge 1.8 \pm 0.9 \,\MeV \\
\Gamma(H_b(1p)\rightarrow\pi^+\pi^- h_b(2p)) &\ge 2.9 \pm 1.5 \,\MeV
\end{aligned}
\]
Certainly these predictions should not be taken for granted unless they were evaluated in an independent manner. Unfortunately, the QCDME has no predictive power for the $H_b(1p)\rightarrow\pi^+\pi^- h_b(np)$ decays since the lack of data on hybrids makes impossible to fix confidently the unknown constants. Furthermore we do not know of any other effective theoretical approach being (successfully) applied to the calculation of these decays. Instead we can only add that the predicted values, although large as compared to the $\Upsilon(5s)\rightarrow\pi^+\pi^- \Upsilon(n_f s)$ widths, may represent a small branching fraction if we rely on constituent quark model estimates for the width of the $1^{--}$, $P-$ wave hybrid states \cite{Idd98}. In these models the hybrid $H_b(1p)$ would dominantly decay to open bottom meson-meson channels, with a width of the order of GeV. This large width might compensate the small $\sin^2\theta$ factor to give a significant contribution to the open bottom meson-meson decays of $\Upsilon(10860)$. The other way around, a thorough independent analysis of these decays, which is completely out of the scope of this letter, could constrain the values of the hybrid width and serve as a stringent test of this kind of models. If these were confirmed, there would be little hope of a direct clean experimental signal of such a broad $H_b(1p)$, or more precisely of the orthogonal combination to~\eqref{mixing} mostly dominated by $H_b(1p)$. This would make our proposal, if correct, the only practical available manner to infer the existence of $H_b(1p)$. Meantime we may only consider the proposed mixing interpretation as a plausible explanation of $\Upsilon(10860)$.

\bigskip

This work has been supported by \foreignlanguage{spanish}{Ministerio de Ciencia, Innovación y Universidades} of Spain and EU Feder under grant FPA2016-77177-C2-1-P and by SEV-2014-0398. R. B. acknowledges the \foreignlanguage{spanish}{Ministerio de Ciencia, Innovación y
Universidades} of Spain for a FPI fellowship.

\bibliography{upsibib}
\end{document}